\documentclass[aps,prl,twocolumn,citeautoscript,superscriptaddress]{revtex4}

\usepackage{amsmath}
\usepackage{amssymb}
\usepackage{graphicx,bm}
\usepackage[caption=false]{subfig}
\usepackage{verbatim}
\usepackage{epstopdf}
\usepackage{dcolumn}
\usepackage{mathtools}
\usepackage{floatrow}
\usepackage{float}
\usepackage{booktabs}
\usepackage[utf8]{inputenc}
\usepackage{array}
\usepackage{makecell}
\usepackage{booktabs}
\usepackage{multirow}
\usepackage{dcolumn}
\usepackage{footnote}

\usepackage{lipsum}

\usepackage{xcolor}
\usepackage{soul}

\UseRawInputEncoding

\begin{document}

\preprint{APS}

\title{Multiple ionization and charge equilibration in slow, multiply charged $\mathrm{Ar^{q+} + Ar}$ collision studied via L-MM Auger-Meitner electron spectroscopy}

\author{Rohit Tyagi}
\affiliation{Department of Physics, Indian Institute of Technology Kanpur, Kanpur - 208016, India}
\author{L. C. Tribedi}
\affiliation{Tata Institute of Fundamental Research, 1 Homi Bhabha Road, Colaba, Mumbai - 400005, India}
\author{M. K. Harbola}
\affiliation{Department of Physics, Indian Institute of Technology Kanpur, Kanpur - 208016, India}
\author{A. H. Kelkar}
\email{akelkar@iitk.ac.in}
\affiliation{Department of Physics, Indian Institute of Technology Kanpur, Kanpur - 208016, India}


\begin{abstract}

We report energy and angle resolved absolute cross section measurements for L-MM Auger-Meitner electron emission following collisions of hundred keV protons and $\mathrm{Ar^{3+/6+}}$ ion with an atomic Ar target. The double differential cross section spectra show distinct contributions from target and projectile Auger–Meitner decay. The projectile emission exhibits the expected Doppler shift for various angles of electron emission, and the measured peak energies are in excellent agreement with kinematic calculations. The energy integrated cross sections show isotropic angular distribution for target and projectile species in their respective rest frames. The Auger-Meitner peak energy for target as well as projectile emission show significant difference in comparison to the characteristic L-MM Auger-Meitner energy peak from atomic Ar. The experimental measurements have been complimented with development of a theoretical model to calculate the transition probabilities corresponding to prominent L-MM Auger-Meitner transitions in neutral and multiply charged Ar atom. Comparison between measured and calculated spectra shows that the measured emission peak at approximately 150 eV originates from Auger–Meitner decay of $\mathrm{Ar^{4+}}$ ions. The peak energies for target and projectile emission are found to be equal, independent of the initial projectile charge state. This indicates that the decay occurs following extensive multiple ionization, charge exchange processes resulting in charge-state equilibration of the collision partners. The results demonstrate that collision-induced electronic rearrangement strongly modifies the Auger–Meitner spectra and provide evidence for an equilibrium target-projectile charge state in low-energy $\mathrm{Ar^{q+} - Ar}$ collisions.

\end{abstract}

\maketitle


\section{Introduction}

In energetic ion-atom collisions with low Z projectile-target pair, inner-shell ionization is predominantly accompanied by relaxation via the Auger-Meitner electron emission channels. The inner-shell vacancy creation occurs in close impact parameter collisions and in case of highly energetic and/or multiply charged projectile ions, it is often accompanied by multiple ionization of the outer shells. Multiple vacancy production modifies the electronic configuration of the collision partners. This results in shift of the electronic energy levels altering the energies and branching ratios of the subsequent Auger-Meitner decay channels. While multiple vacancies in the M-shell weakly influences K-shell transitions, it can substantially modify the energies of L-shell Auger-Meitner transitions, making Auger–Meitner spectroscopy a sensitive probe of collision-induced electronic rearrangement. Collisions between symmetric target atom - ion pairs ($\mathrm{X^{q+} + X}$) have been proposed to proceed through formation of transient quasi-molecular states leading to inner-shell vacancy production \cite{DEGROOT1987159, RShanker1982, Lichten, FanoLi, GARCIA, Barat1972} with direct excitation by the projectile nucleus having relatively feeble contribution. The larger interaction volume associated with the quasi-molecular system also results in significantly enhanced core-shell ionization cross sections in comparison to collisions involving structureless projectiles such as protons or electrons. A theoretical framework had also been proposed by Eichler \textit{et al.} \cite{Eichler1,Eichler2} to incorporate the influence of the projectile charge state on the electronic rearrangement of target electron energy levels during collision. 

The $\mathrm{Ar^{1+} + Ar}$ collision system has subsequently been studied through measurements of K-shell x-ray production cross sections and x-ray spectra \cite{Tawara,PHWoerlee_1981}. These studies showed that the K-shell x-ray yield is largely independent of the initial projectile charge state, suggesting rapid charge redistribution during the collision. Since the Auger–Meitner process is complementary to x-ray emission and is considerably more sensitive to the electronic configuration of the ionized system, measurements of AM electron spectra can provide direct information on multiple ionization, charge exchange, and charge-state equilibration. Numerous investigation have also studied Ar L-MM and Ne K-LL Auger-Meitner spectra, in collision with low energy $\mathrm{Ar^{q+}}$ and $\mathrm{Ne^{q+}}$ ($\mathrm{q\le 2}$), respectively \cite{Morgan,Everhart,StolterfohtNe,Rudd2,Rudd1,Schiwietz}. The measurements were performed in a limited angular range and revealed the characteristic features of target and projectile Auger-Meitner emissions. However, the effects of multiple ionization and electron energy level rearrangements were not substantial due to low charge state of the projectile ion.

In the present study, we report absolute double differential cross-section measurements of L-MM Auger-Meitner electron emission in collisions of 600 keV $\mathrm{Ar^{3+}}$ and $\mathrm{Ar^{6+}}$ ions with atomic Ar. Owing to low velocity of the projectile ion ($\mathrm{v_p \sim 0.7}$ a.u.), the measured electron spectra contain contributions from both the target and the projectile, with the latter showing a Doppler shift in energy in the laboratory frame. We observed a substantial shift in the Auger-Meitner peak energies for $\mathrm{Ar^{3+/6+} + Ar}$ collision system from the characteristic Auger-Meitner peak energy for Ar atom. We shall also discuss a theoretical model based on Hartree-Fock Self Consistent Field approximation and Slater orbitals, developed to simulate the L-MM Auger-Meitner spectra for multiply charged Ar ions. The measured and calculated spectra are compared to highlight the role of collision-induced electronic rearrangement in Auger–Meitner decay and reveal the importance of multiple vacancy creation in slow, highly charged ion–atom collisions.

\section{Experiment}
The L-MM Auger-Meitner electron emission cross sections were measured using the electron spectrometer setup at the 300 keV ECR ion accelerator facility (ECRIA), TIFR, Mumbai \cite{Agnihotri2011}. $\mathrm{H^{+}}$, $\mathrm{Ar^{3+}}$, and $\mathrm{Ar^{6+}}$ projectile ion beams were obtained from the 14.5 GHz ECR ion source floated at 30 keV. The ion beams were further accelerated, after mass selection, to the desired energy by raising the 300 kV high voltage deck to appropriate potential. Post acceleration a switching magnet directed the ion beam to the $\mathrm{30^o}$ beamline. The ion beam was collimated and focused using an electrostatic quadrupole lens triplet and a pair of four jaw slits before entering the main scattering chamber. the ion-target gas collision experiments were performed under static target gas pressure conditions. The main scattering chamber was flooded with the target gas and the absolute number density of the target gas atoms was calculated by measuring the gas pressure using a capacitance manometer. The base pressure of the main scattering chamber was $\mathrm{\le 5\times 10^{-8}}$ mbar whereas the target gas pressure was kept below $\mathrm{5\times 10^{-5}}$ mbar during the experiments thereby maintaining single collision conditions. The ejected secondary electrons were energy analyzed using an electrostatic hemispherical analyzer and detected by a channel electron multiplier. The analyzer was mounted on a rotary platform to measure electron emission at various angles ($\mathrm{20^o - 150^o}$) in steps of $\mathrm{15^o}$ relative to the projectile ion beam direction. The energy resolution of the electron spectrometer is $\mathrm{\sim 6 \%}$. Details of the electron spectrometer and the ECRIA facility at TIFR have been described earlier in detail \cite{Misra2009, Agnihotri2011}. 

The Absolute double differential electron emission cross section (DDCS) is calculated using the following relation:
\begin{equation}
    \mathrm{
        \frac{d^2\sigma}{d\Omega d\epsilon} = \frac{\left( \frac{n_e}{N_p\Delta\epsilon} - \frac{n_b}{N_p\Delta\epsilon} \right)}{P_t\epsilon_{el}(l\Omega)_{eff}}
    }
\end{equation}

Here, $\mathrm{n_e}$ and $\mathrm{n_b}$ are detected electron counts from the target gas and background respectively, $\mathrm{N_p}$ is the number of projectile ions, $\mathrm{P_t}$ is the target gas pressure, $\mathrm{\Delta\epsilon}$ is the energy resolution of the emitted electrons, $\mathrm{\epsilon_{el}}$ is the detection efficiency and $\mathrm{(l\Omega)_{eff}}$ is the effective solid angle path length \cite{Misra2009}. Uncertainty in the measurement of the target gas pressure ($\mathrm{\sim 10\%}$) is the dominant source of error in the calculation of absolute DDCS, other than the statistical error. The over all uncertainty in absolute DDCS is estimated to be $\mathrm{\sim 15\%}$.

\section{Result and discussion}

\begin{figure*}
    \centering
    \includegraphics[width = \textwidth]{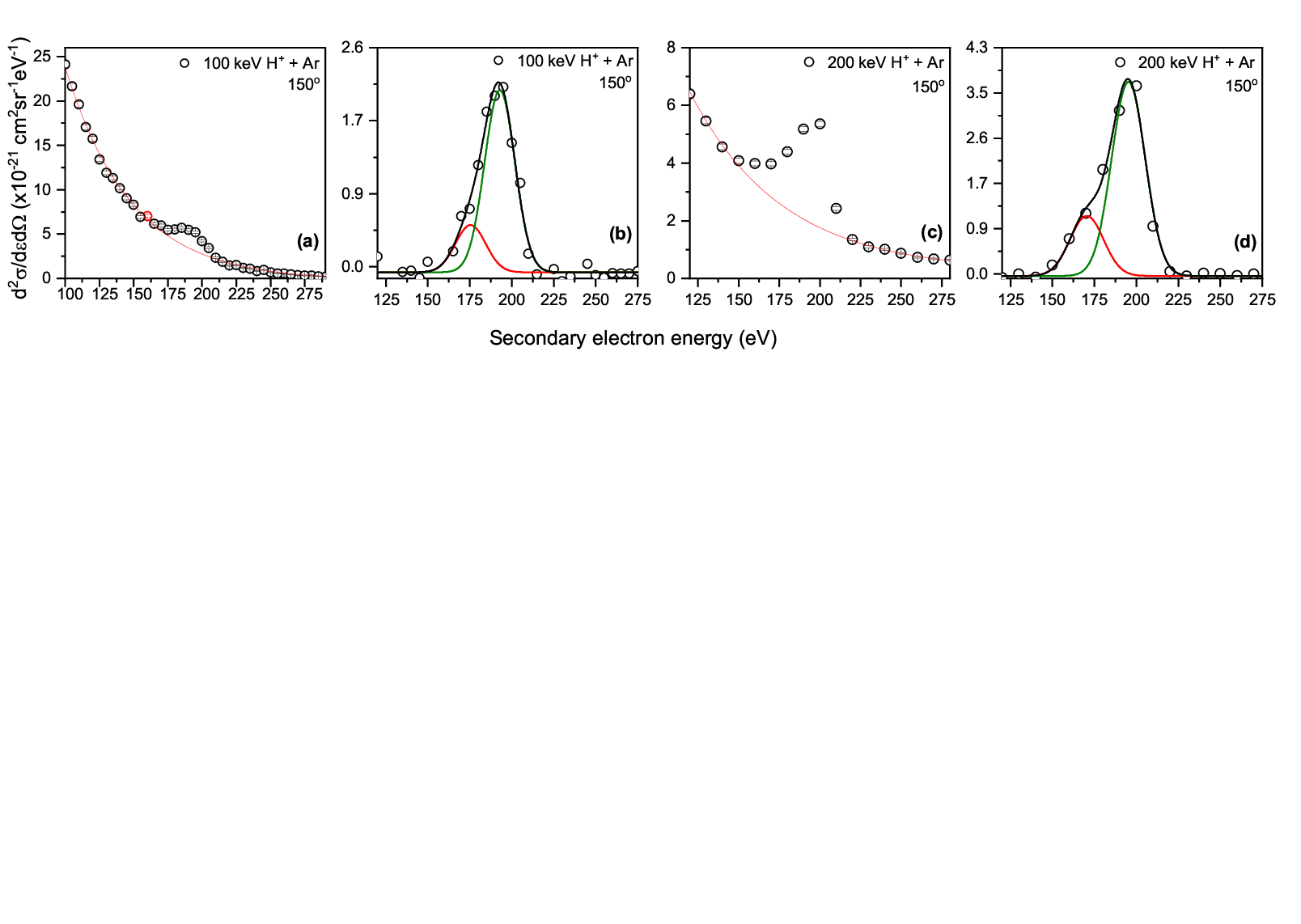}
    \vspace{-8cm}
    \caption{Absolute DDCS spectra at 150$^o$, for Ar in collision with (a and b) 100 keV proton ion beam  and (c and d) 200 keV proton ion beam. The DDCS in panel (a) and (c) show the full DDCS spectrum (including the Coulomb continuum background). The solid (red) curve in (a) and (c) is a polynomial background fit to estimate continuum electron background in the Auger - Meitner energy region. Panel (b) and (d) show the background subtracted Ar L-MM Auger - Meitner peaks. Solid (Red and Green) curves in the panels (b and d) show Gaussian fits with centroid energy equal to 190 eV and 175 eV respectively. The solid - black curve shows the overall fit to the measured data. The error bars represent the statistical error in the measured data. Absolute error in DDCS measurement is $\sim$15$\%$. }
    \label{Fig1}
\end{figure*}

\begin{figure}
    \centering
    \includegraphics[width = \textwidth]{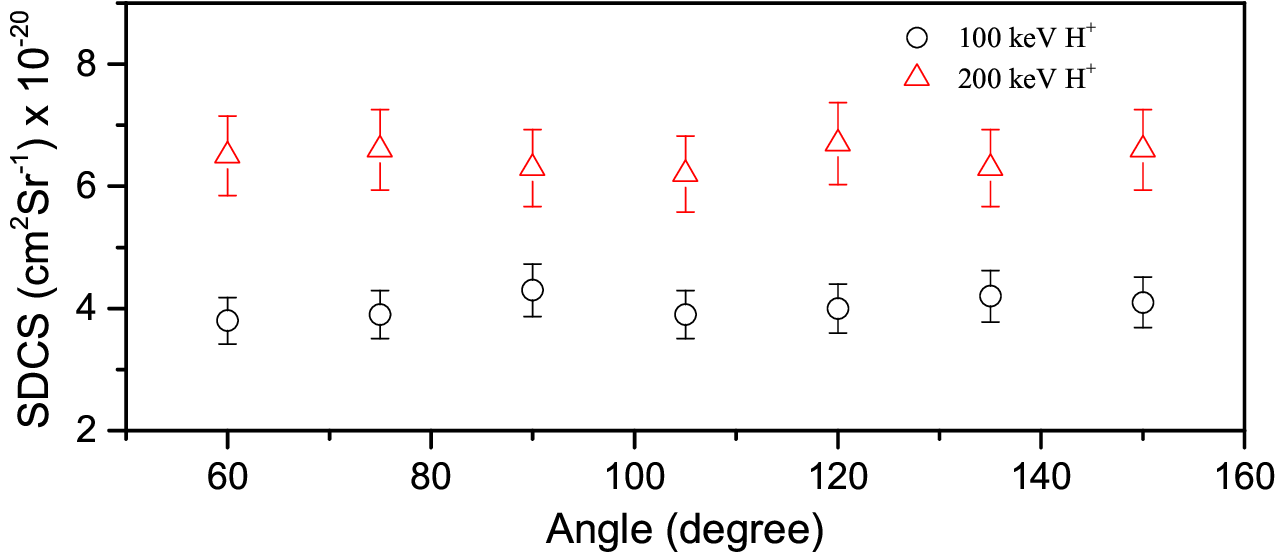}
    \caption{Angular SDCS for 100 keV (black circle) and 200 keV (Red triangle) $\mathrm{H^+ + Ar}$ collision. The error bars represent absolute error ($\sim$ 15$\%$) in the cross section measurement.}
    \label{Fig2}
\end{figure}

\begin{figure*}
    \centering
    \includegraphics[width = \textwidth]{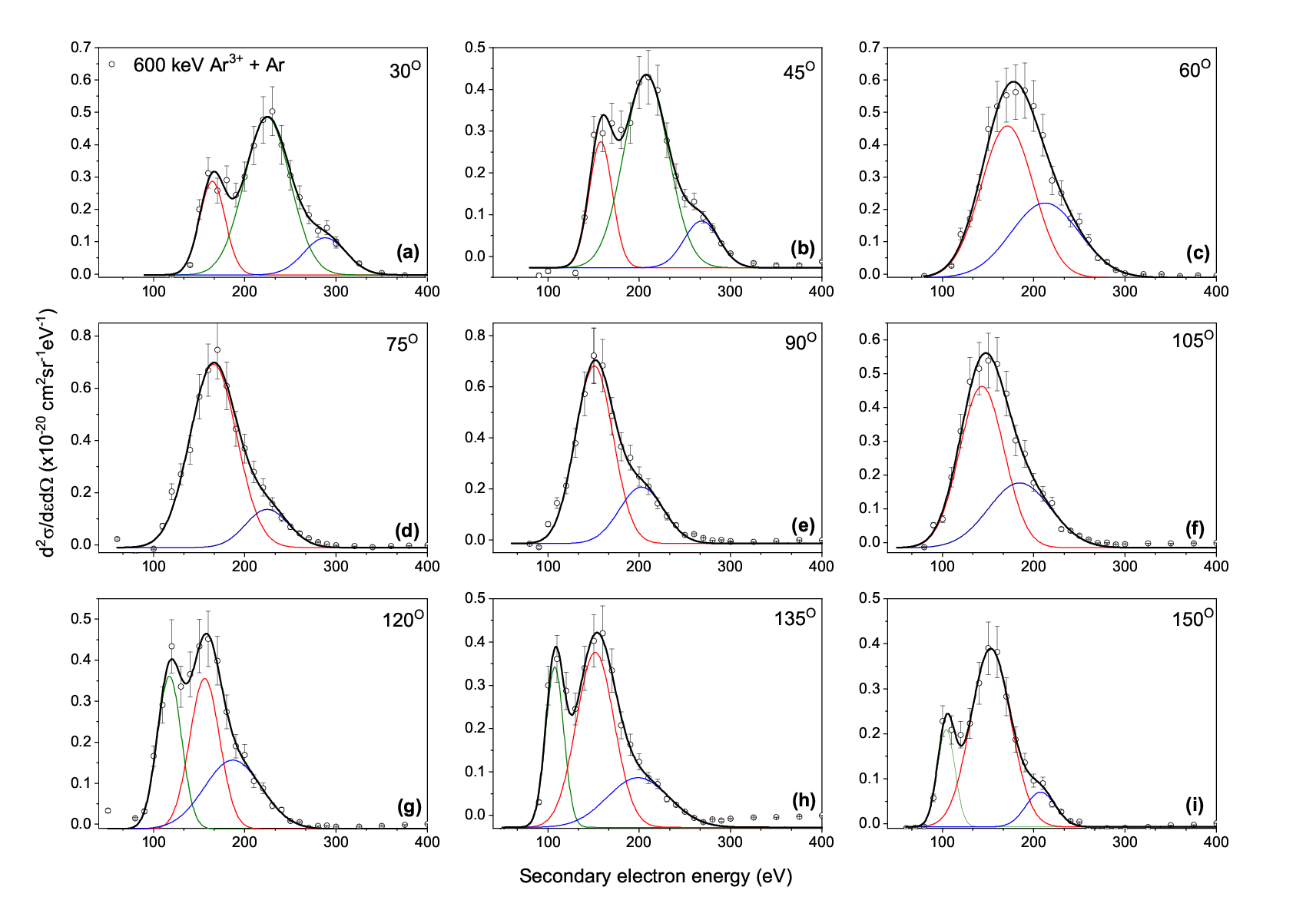}
    \caption{Coulomb continuum background subtracted electron DDCS for 600 keV $\mathrm{Ar^{3+} + Ar}$ collision. Solid colored curves (red, green and blue) show Gaussian peak fits and solid black curve shows the combined fit to the spectra in each panel. Centroid energy of the peak fits (red and green) are given in table \ref{Table1}. The error bars represent the statistical error in the measured data. Absolute error in DDCS measurement is $\sim$15$\%$. }
    \label{Fig3}
\end{figure*}

\begin{figure*}
    \centering
    \includegraphics[width = \textwidth]{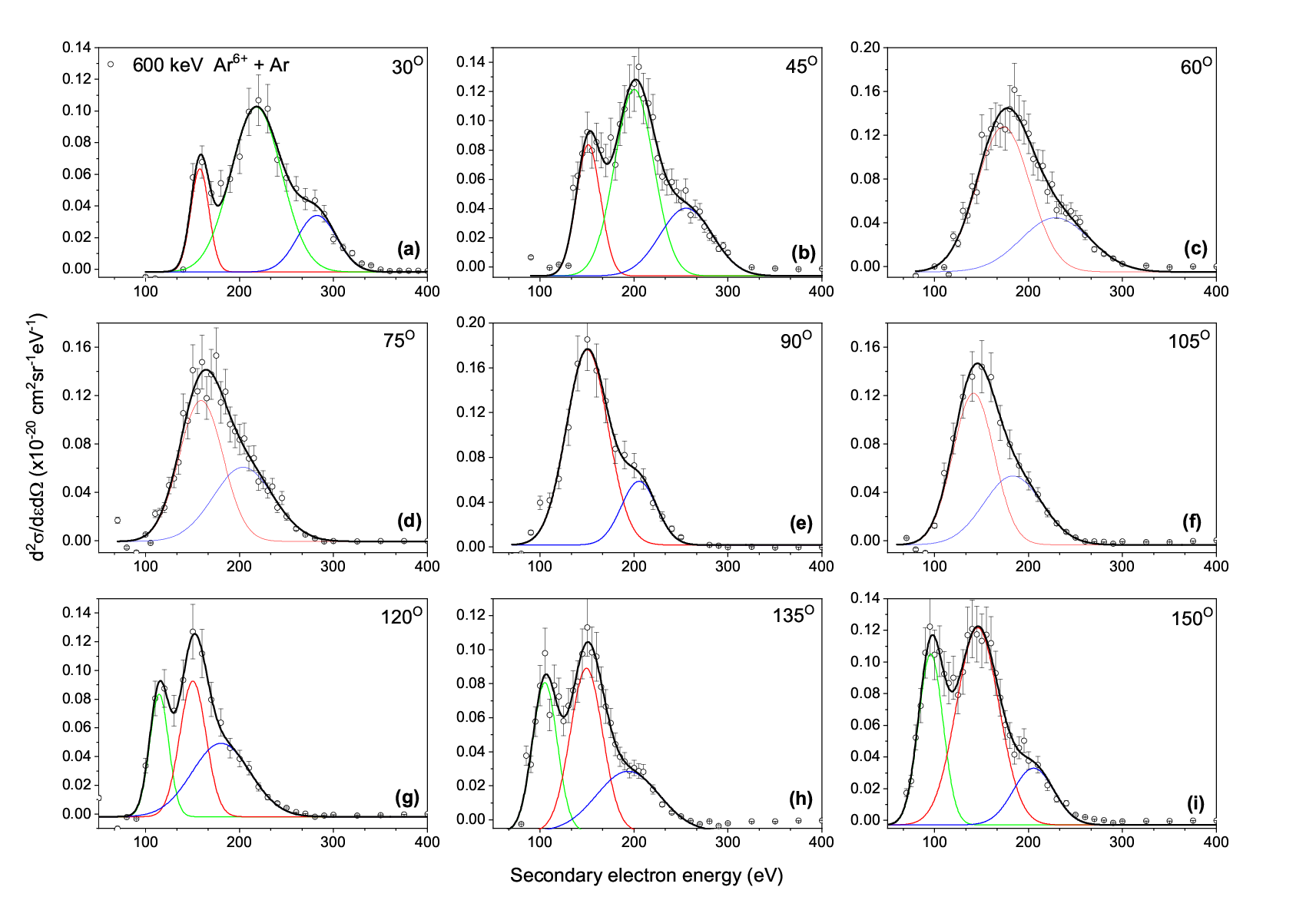}
    \caption{Coulomb continuum background subtracted electron DDCS for 600 keV $\mathrm{Ar^{6+} + Ar}$ collision. Solid colored curves (red, green and blue) show Gaussian peak fits and solid black curve shows the combined fit to the spectra in each panel. Centroid energy of the peak fits (red and green) are given in table \ref{Table1}. The error bars represent the statistical error in the measured data. Absolute error in DDCS measurement is $\sim$15$\%$.}
    \label{Fig4}
\end{figure*}

\begin{figure}
    \centering
    \includegraphics[width = 1.8\textwidth]{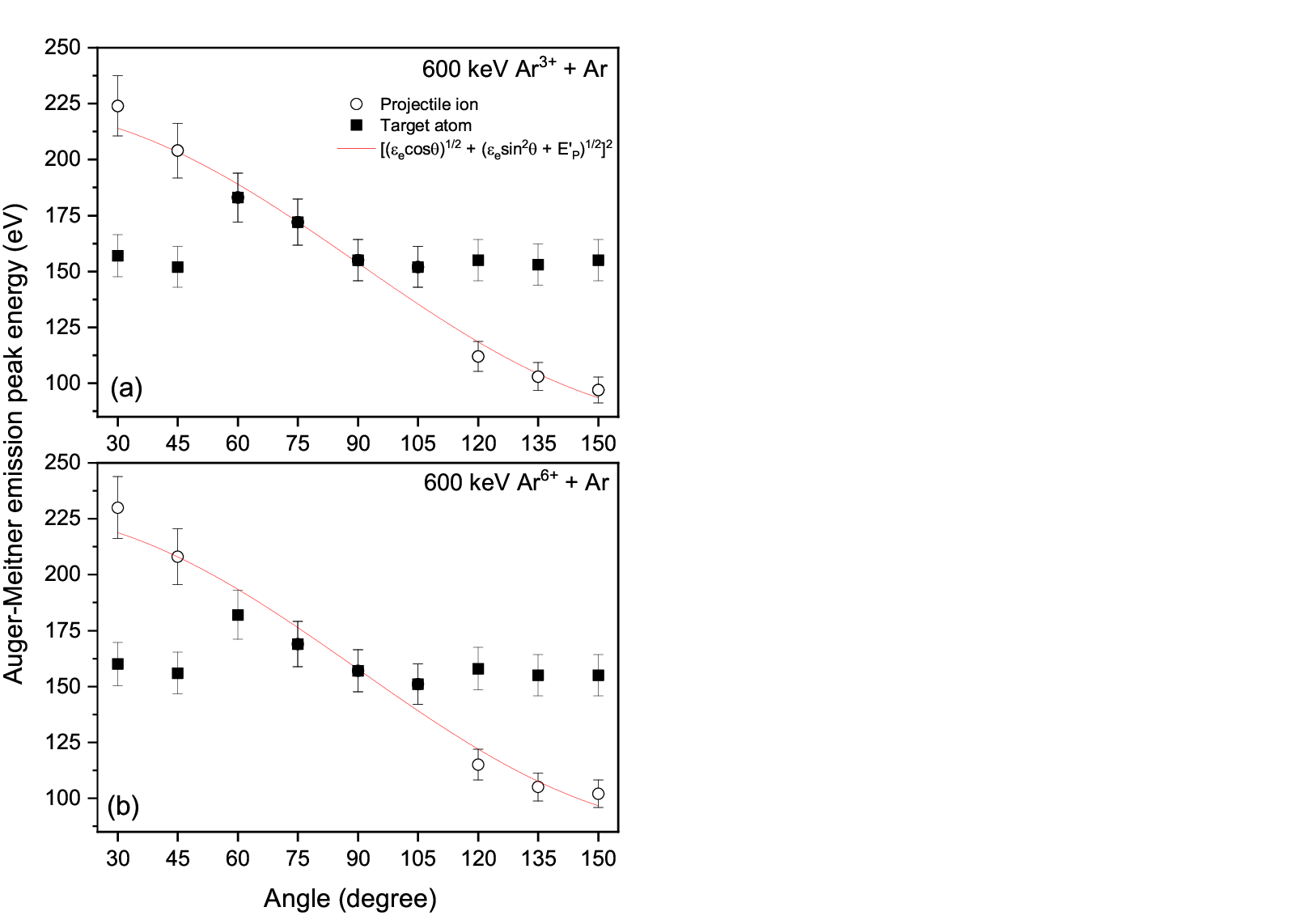}
    \caption{Target (filled squares) and projectile (hollow circles) Auger-Meitner peak energies (as tabulated in table \ref{Table1}) as a function of angle of electron emission measured with respect to the projectile ion beam direction. (a) $\mathrm{Ar^{3+} + Ar}$ (b) $\mathrm{Ar^{6+} + Ar}$. The red solid curve in both (a) and (b) shows the calculated projectile peak energy ($\mathrm{E_P}$) in equation \ref{Eq2}. }
    \label{Fig5}
\end{figure}

\begin{figure}
    \centering
    \includegraphics[width = 1.8\textwidth]{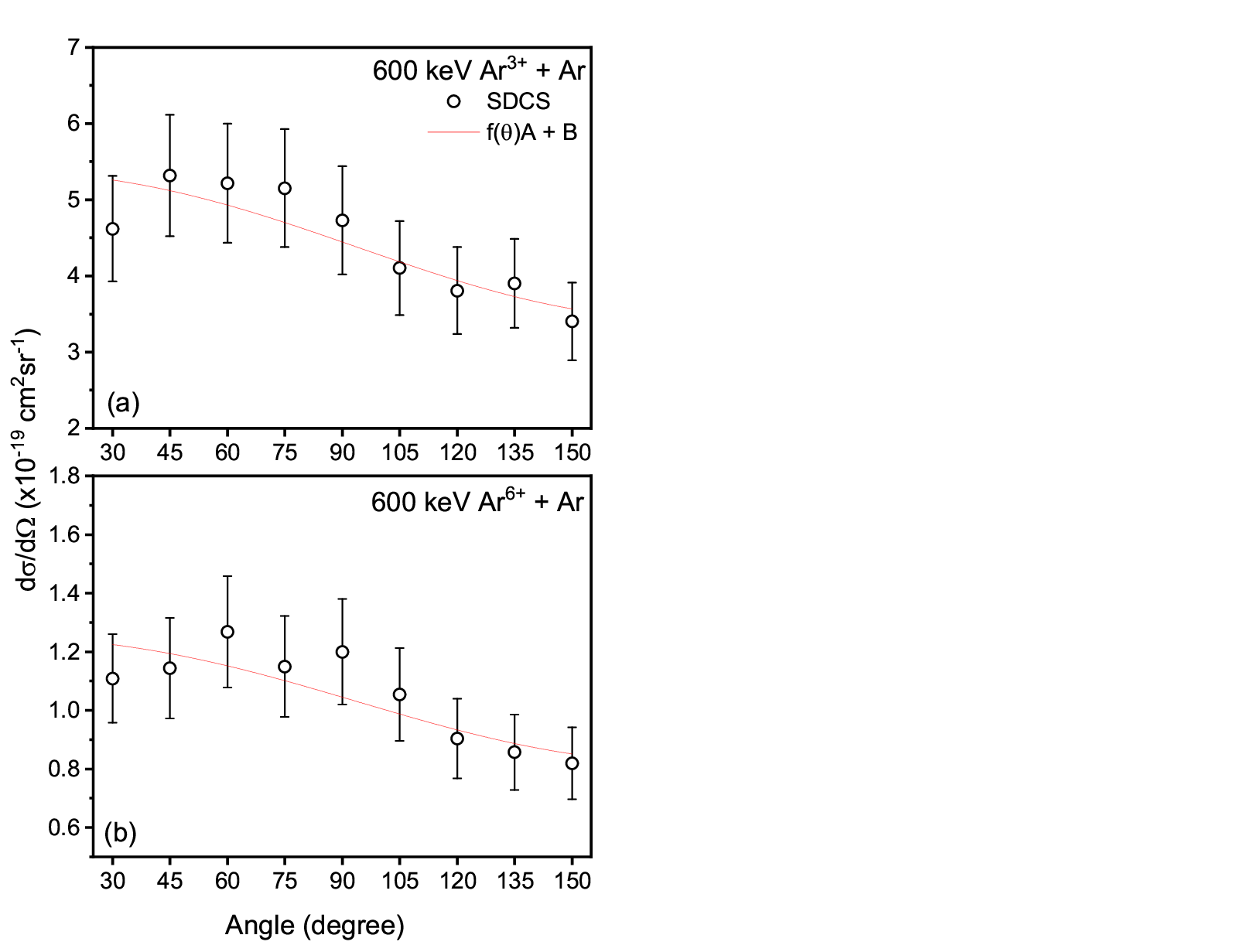}
    \caption{Total angular SDCS (obtained by energy integrating the DDCS spectra in figures \ref{Fig3} and \ref{Fig4}).(a) $\mathrm{Ar^{3+} + Ar}$ (b) $\mathrm{Ar^{6+} + Ar}$. The red solid curve in both (a) and (b) shows the optimized fit of equation \ref{Eq3} considering the isotropic nature of target SDCS $\mathrm{\left(B = \frac{d\sigma_t}{d\Omega}\right)}$ and projectile SDCS $\mathrm{\left(A = \frac{d\sigma'_p}{d\Omega'}\right)}$. The function $\mathrm{f(\theta)}$ represents the Doppler correction in projectile rest frame (see equation \ref{Eq3}). }
    \label{Fig6}
\end{figure}

Energy and angle differential absolute cross sections of L-MM Auger-Meitner electron emission were measured for an Ar atomic target in collisions with proton and multiply charged $\mathrm{Ar^{3+}}$ and $\mathrm{Ar^{6+}}$ projectile ions. 

\subsection{ H$^+$ + Ar collision}

The measurements with proton beam were performed at two different beam energies of 100 keV and 200 keV. We measured the energy spectrum of the emitted secondary electrons in the range 110 eV to 300 eV. This energy range covers the L-MM Auger-Meitner electron energy for neutral Ar atom. The electron spectra were measured at various angles ranging from $\mathrm{30^o - 150^o}$ with respect to the ion beam direction. In figure \ref{Fig1}(a) and (c) we have shown the measured electron spectrum at $\mathrm{150^o}$ for 100 keV and 200 keV proton beams, respectively. The electron spectra show the L-MM Auger-Meitner peak along with the Coulomb continuum background.  The total L-MM Auger-Meitner contribution is further obtained by subtracting the Coulomb continuum background in the L-MM Auger region. For this purpose we simulate the Coulomb background as an appropriate polynomial function in the L-MM peak region and subtract it from the measured DDCS spectrum \cite{Kelkar2020, Kelkar2022}. This interpolated continuum background is shown as solid curve in figure \ref{Fig1}(a) and (c) and the subtracted L-MM Auger-Meitner peaks are shown in figure \ref{Fig1}(b) and (d).

The Auger-Meitner energy spectra of Ar in collisions with proton and other heavy ion projectiles have been studied earlier \cite{volz, Rudd1, Stolte1974}. Volz et al. \cite{volz} measured high resolution Auger-Meitner energy spectrum in 100 - 300 keV $\mathrm{H^+}$ collision with an Ar target. With a spectrometer resolution of 0.9 eV, the authors identified 27 peaks including satellite peaks in the Auger-Meitner region in the energy range 150 eV to 210 eV.  The dominant peaks in the 200 eV - 205 eV range were attributed to $\mathrm{L_{23}-M^2_{23} (^1S, ^1P, ^1D)}$ transitions. The spectra also showed dominant peaks in 175 eV - 190 eV range, resulting from $\mathrm{L_{23}-M^2_1 (^1S)}$ and $\mathrm{L_{23}-M_1M_{23} (^1P, ^3P)}$ transitions. Owing to the relatively poor resolution of our spectrometer, we observe the Auger-Meitner peak as a broad hump in the given energy range. In order to compare our measured Auger-Meitner spectrum with the high resolution results of Volz et al. \cite{volz} we regenerated the high resolution spectra considering the peak energies identified earlier and experimental resolution reported by the authors. Further, we convoluted each peak with the resolution of our spectrometer ($\mathrm{\sim 6\%}$) and used the branching ratios reported by volz et al. \cite{volz} as weight factor to generate the overall distribution. The generated Auger-Meitner distribution has a maxima close to 197 eV with a low energy tail extending down to 150 eV. This is in excellent agreement with our measured spectra (see figure \ref{Fig1}(b) and (d)) having a peak at 194 eV. The measured spectra also show a low energy  tail with a weak shoulder at $\sim$ 175 eV. The measured spectra were fitted with two Gaussian curves (shown with solid lines in figure \ref{Fig1}(b) and (d)) to identify the peak energy. 

The cumulative Auger-Meitner yield (from all possible transition states) at a given angle of electron emission is calculated by integrating the Auger-Meitner peak over the entire energy range. This is termed as the single differential cross section (SDCS). In figure \ref{Fig2} we have shown the SDCS angular distribution. As expected \cite{volz} the angular distribution is isotropic within experimental uncertainty. The isotropic nature of the SDCS angular distribution allows us to calculate the total Auger-Meitner electron emission cross section by taking the mean SDCS from the angular distribution plot (see figure \ref{Fig2}) and integrating it over 4$\pi$ solid angle. The measured absolute total cross section (TCS) for Ar L-MM Auger-Meitner emission are $\mathrm{5\times10^{-19}}$ $\mathrm{cm^2}$ and $\mathrm{8.5\times10^{-19}}$ $\mathrm{cm^2}$ for 100 keV and 200 keV proton beams, respectively. 

\begin{table}
	\caption{L-MM Auger-Meitner peak energies (obtained from multi-peak fit to the DDCS data shown in figures \ref{Fig3} and \ref{Fig4}) at various angles of electron emission with respect to the projectile ion beam direction. $\mathrm{E_{t1}}$ and $\mathrm{E_{P1}}$ correspond to target and projectile peaks for $\mathrm{Ar^{3+} + Ar}$ collision system, respectively. $\mathrm{E_{t2}}$ and $\mathrm{E_{P2}}$ correspond to target and projectile peaks for $\mathrm{Ar^{6+} + Ar}$ collision system, respectively.}      
	\begin{tabular}{|p{0.16\textwidth}|p{0.16\textwidth}|p{0.16\textwidth}|p{0.16\textwidth}|p{0.16\textwidth}|}
			\hline
        Angle (degree) & $\mathrm{E_{t1}}$ (eV) & $\mathrm{E_{P1}}$ (eV) & $\mathrm{E_{t2}}$ (eV) & $\mathrm{E_{P2}}$ (eV)
			 \\
			\hline
			30    & 160   & 228    & 157   & 216    \\ \hline
			45    & 157   & 207   & 151   & 200    \\ \hline
			60    & 178   & 178    & 150   & 186    \\ \hline
			75    & 167   & 167    & 164   & 164    \\ \hline
			90    & 152   & 152    & 149   & 148    \\ \hline
			105   & 147   & 147    & 145   & 145    \\ \hline
			120   & 157   & 117    & 152   & 113    \\ \hline
			135   & 154   & 110    & 150   & 104    \\ \hline
			150   & 153   & 107    & 150   & 98     \\ 
			\hline
	\end{tabular}\label{Table1}
\end{table}

\subsection{ Ar$^{3+/6+}$ + Ar collision}

Unlike protons, inner shell ionization and subsequent Auger-Meitner decay following collision with multiply charged heavy ion projectiles like $\mathrm{Ar^{q+}}$ is predominantly accompanied by multiple vacancy creation. The electronic state distribution of the target ion gets modified significantly compared to the ground state configuration. This rearrangement results in shift in the electronic energy levels thereby modifying the branching ratio of the transition states and the resultant Auger-Meitner spectra. At collision energies in the range of few hundred keV, electron capture processes also play a significant contribution toward inner shell vacancy creation. In the following we present a detailed investigation of Ar L-MM Auger-Meitner yield following collision of 600 keV $\mathrm{Ar^{3+}}$ and $\mathrm{Ar^{6+}}$ projectile ions with Ar atomic target. The $\mathrm{Ar^{3+}}$ and $\mathrm{Ar^{6+}}$ ion beams were generated using a ECR ion source and their initial electron state population distribution may have significant contribution of excited states. However, the ion beam travels close to 5 meters before entering the scattering chamber and takes few microseconds to cover this distance. This travel time is sufficient for all excited state ions to relax to the ground state. Therefore, the electronic configuration of the ion beams is $\mathrm{Ar^{3+} (1s^22s^22p^63s^23p^3)}$ and $\mathrm{Ar^{6+} (1s^22s^22p^63s^2)}$. 

In figure \ref{Fig3} we have shown the DDCS electron spectra for 600 keV $\mathrm{Ar^{3+}}$ + Ar collision measured at various emission angles. The Coulomb continuum background has been subtracted in each of the DDCS plots to show only the L-MM Auger-Meitner electron emission peak. The DDCS spectra in figure \ref{Fig3} show two dominant broad peaks covering a wider energy range at forward and backward angles ($\mathrm{30^o, 45^o, 120^o, 135^o}$ and $\mathrm{150^o}$) whereas this changes to a single broad peak at angles close to $\mathrm{90^o}$ ($\mathrm{60^o, 75^o, 90^o}$ and $\mathrm{105^o}$). An additional peak is also identified after peak fitting on the high energy tail of the spectra. Similar DDCS spectra for collisions with 600 keV $\mathrm{Ar^{6+}}$ projectile ions are also shown in figure \ref{Fig4}. Using multi-peak fitting analysis we have extracted the peak energies and tabulated the same in table \ref{Table1}. A careful consideration of the tabulated energy peak values and the DDCS spectra reveals that the electron DDCS spectra in figure \ref{Fig3} consists of two main Auger-Meitner peaks. One peak ($\mathrm{E_{t1}}$) has a fixed centroid energy value $\mathrm{\sim 150-160}$ eV. The centroid of the second peak ($\mathrm{E_{P1}}$) varies as a function of angle of emission going from 228 eV at $\mathrm{30^o}$ to 107 eV at $\mathrm{150^o}$. The characteristic Auger-Meitner peak energy from a stationary atom/ion is independent of the angle of emission in the laboratory frame of measurement (cf figure \ref{Fig1}). Therefore, stationary peak centered at energy $\mathrm{E_{t1}}$ is assigned to target L-MM Auger-Meitner decay. On the other hand, the Auger-Meitner peak energy corresponding to a moving atom/ion would be Doppler shifted in the laboratory frame of measurement, where the Doppler shift in energy is a function of angle of emission. Therefore, the peak $\mathrm{E_{P1}}$, varying with angle of emission, could be attributed to the Auger-Meitner electron emission from $\mathrm{Ar^{3+}}$ (and $\mathrm{Ar^{6+}}$) projectile ion. The Doppler shift in the energy of the Auger-Meitner electron from a moving projectile is given as \cite{StolterfohtNe, stolterfoht1987}:

\begin{equation}
\mathrm{
E_P = \left[\sqrt{\epsilon_e}\cos{\theta} \pm \sqrt{\epsilon_e\sin^2{\theta} + E'_P}\right]^2
}
\label{Eq2}
\end{equation}

Here, $\mathrm{E_P}$ and $\mathrm{E'_P}$ are the energies of Auger-Meitner electron in the laboratory frame and projectile frame, respectively. $\mathrm{\theta}$ is the angle of emission and $\mathrm{\epsilon_e}$ is the energy of an electron moving with the speed of the projectile ion. For 600 keV $\mathrm{Ar^{3+}}$ (and $\mathrm{Ar^{6+}}$) projectile ion the corresponding $\mathrm{\epsilon_e = 8.2}$ eV. Using $\mathrm{E'_P = 150}$ eV (same as $\mathrm{E_{t1}}$ in table \ref{Table1}), we have calculated the Doppler shifted projectile Auger-Meitner peak energy ($\mathrm{E^{Cal}_{P1}}$) for various angles of measurement. In figure \ref{Fig5} we have plotted the measured Auger-Meitner peak energy corresponding to electron emission at different angles from stationary Ar target (filled circles) as well as (a) $\mathrm{Ar^{3+}}$  and (b) $\mathrm{Ar^{6+}}$ projectile ion (open circles). The calculated projectile Auger-Meitner energy values ($\mathrm{E^{Cal}_{P1}}$) are also shown as the solid curve in the same plot. It is evident from the plot (see figure \ref{Fig5})that the calculated energy values ($\mathrm{E^{Cal}_{P1}}$) are in excellent agreement with the measured peak energy ($\mathrm{E_{P1}}$).

The DDCS plots for each angle of emission (figure \ref{Fig3} and \ref{Fig4}) can be further integrated over the Auger-Meitner energy range to obtain the angular SDCS. The energy range for integration contains the SDCS of target as well as projectile ions. The SDCS values for target and projectile ions can be separately obtained using the Gaussian fits in figure \ref{Fig3} and \ref{Fig4}. However, due to the varied over lap of the two Auger-Meitner peaks and their indistinguishably for angles close to $\mathrm{90^o}$, we have used an alternate procedure to extract target and projectile SDCS. 

The total SDCS, target and projectile ion combined, is given as \cite{StolterfohtNe, stolterfoht1987}:

\begin{equation}
\mathrm{
	\frac{d\sigma}{d\Omega} = \left(\frac{E_P}{E'_P}\right)\sqrt{\left(1-\frac{\epsilon_e}{E'_P}\sin^2\theta\right)}\frac{d\sigma'_p}{d\Omega'}+\frac{d\sigma_t}{d\Omega}
    }\label{Eq3}
\end{equation}

Here, $\mathrm{\frac{d\sigma}{d\Omega}}$ is the measured SDCS (energy integrated DDCS over the entire Auger-Meitner energy range covering both,  target and projectile peaks). $\mathrm{\frac{d\sigma_P'}{d\Omega'}}$ is the SDCS corresponding to Auger-Meitner emission from the projectile ion in its rest frame and $\mathrm{\frac{d\sigma_t}{d\Omega}}$ is the SDCS corresponding to Auger-Meitner emission from the target atom in the laboratory frame.

In figure \ref{Fig6} we have plotted the measured total angular SDCS $\mathrm{\frac{d\sigma}{d\Omega}}$, (open circles) for all angles of measurement. The measured SDCS have been fitted with the expression in equation \ref{Eq3} (Solid line in figure \ref{Fig6}). For fitting, we have considered the target and projectile SDCS (in their respective rest frames) to be isotropic. It is evident that the expression in equation \ref{Eq3} fits the measured SDCS data very well and reproduces the over all angular distribution. Further, the values of projectile ion SDCS $\mathrm{\left(\frac{d\sigma_P'}{d\Omega'}\right)}$ and target SDCS $\mathrm{\left(\frac{d\sigma_t}{d\Omega}\right)}$ were obtained from the fit. The target and projectile ion SDCS values were further multiplied by $\mathrm{4\pi}$ (as is the case for isotropic angular distribution) to calculate the total L-MM Auger-Meitner electron emission cross section for Ar target and $\mathrm{Ar^{3+}}$ and $\mathrm{Ar^{6+}}$ projectile ion. The calculated total cross sections are given in table \ref{Table2}.

\begin{table}
\caption{Estimated total absolute cross section (TCS) for L-MM Auger-Meitner electron emission for Ar target ($\mathrm{\sigma_t}$) and $\mathrm{Ar^{q+}}$ (q = 3, 6) projectile ion ($\mathrm{\sigma_P}$).}
\begin{tabular}{ |p{0.15\textwidth}|p{0.4\textwidth}|p{0.4\textwidth}|  }
\hline
Projectile ion & $\mathrm{\sigma_t}$ in laboratory frame ($\mathrm{\times 10^{-18}}$ $\mathrm{cm^2}$) & $\mathrm{\sigma_P}$ in projectile ion rest frame ($\mathrm{\times 10^{-18}}$ $\mathrm{cm^2}$)
\\ \hline
$\mathrm{Ar^{3+}}$  & $\mathrm{2.7 \pm 0.5}$  & $\mathrm{2.5 \pm 0.5}$\\ \hline
$\mathrm{Ar^{6+}}$  & $\mathrm{0.7 \pm 0.5}$  & $\mathrm{0.5 \pm 0.5}$\\ \hline
\end{tabular}\label{Table2}
\end{table} 

In the preceding discussion, we have presented a detailed analysis of the observed Auger-Meitner electron emission spectra in $\mathrm{Ar^{q+}}$ + Ar collision system. Characteristic features of the spectra including angular distribution are explained on the basis of Auger-Meitner emission from the target and projectile species, both. Kinematic considerations permit us to calculate the Auger-Meitner peak energy in the projectile rest frame. The energy peak values are similar ($\mathrm{\sim 150}$ eV) for target (atom) and projectile (ion) in their respective rest frames. This peak energy value is far separated from the expected Auger-Meitner peak energy ($\mathrm{\sim 190 - 200}$ eV) from an atomic Ar target (see figure \ref{Fig1}). However, the Auger-Meitner peak from a multiply charged $\mathrm{Ar^{q+}}$ ion would appear at lower energy (compared to atomic Ar) due to increased binding energy of the outer valence electrons. In addition, it is well known that collision with slow, multiply charged projectile ion leads to multiple ionization/excitation of the target atom due to strong Coulomb interaction and large dwell time. Therefore, the observed Auger-Meitner spectra with peak energy $\mathrm{\sim 150}$ eV would correspond to a multiply charged target - projectile combination. The similarity in the Auger-Meitner peak energy for $\mathrm{Ar^{3+}}$ and $\mathrm{Ar^{6+}}$ projectile ions, further suggests that, for the chosen collision system, the Auger-Meitner process follows multiple vacancy creation in the target atom and charge equilibration of the target $-$ projectile ion pair. 

The equilibrium charge state of the target $-$ projectile ion pair can be ascertained by calculating the Auger-Meitner transition energies and corresponding transition probabilities for multiply charged Ar ions. In the following section we present a theoretical framework to calculate the transition probabilities of prominent Auger-Meitner decay transitions for atomic Ar and $\mathrm{Ar^{q+}}$ ($\mathrm{q =2 - 4}$) ions.

\subsection{Transition probablity calculation}
Auger-Meitner process is a result of Coulomb interaction between two electrons in singly ionized (core shell ionization) atom. In this process an initial vacancy state ($\mathrm{n_1, l_1}$) decays into a final state with two vacancies ($\mathrm{n_2, l_2}$),($\mathrm{n_3, l_3}$). Probability of this transition (in atomic units) is given by the Fermi golden rule  

\begin{equation}
\mathrm{
   t_{i \to f} = \Big |\int \Phi^*_f V(r)\Phi_i d\tau \Big |^2 \rho(E_f)
    }
   \label{Eq4}
\end{equation}

\[\mathrm{
    \Phi_f = \psi_{M_b}(1)\psi_{M_c}(2)}\]
    \[\mathrm{\Phi_i = \psi_{L_a}(1)\psi_\infty(2) 
}\]


Here $\mathrm{\rho(E_f)}$ is the density of states for the energy $\mathrm{E_f}$ and $\mathrm{\Phi_f}$ is the final state wavefunction with hole (1) occupying $\mathrm{M_b}$ shell and electron (2) in $\mathrm{M_c}$ shell. Similarly, $\mathrm{\Phi_i}$ is the initial state wavefunction with hole (1) in $\mathrm{L_a}$ shell and hole (2) in the continuum. The interaction potential $\mathrm{V(r) = \frac{1}{r_{12}}}$.

Using the spherical harmonics, we can write

\begin{equation}
\mathrm{
      \frac{1}{r_{12}} = \frac{4\pi}{2\lambda+1}\sum_\lambda (-1)^{m_\lambda}\nu(r_1,r_2)Y_{\lambda m_\lambda}(\theta_1\phi_1)Y_{\lambda -m_\lambda}(\theta_2\phi_2)}
\end{equation}

\[
\mathrm{
   \nu(r_1,r_2)= 
\begin{dcases}
     \frac{r_1^\lambda}{r_2^{\lambda+1}} ,& \text{if } r_2\geq r_1\\
     \frac{r_2^\lambda}{r_1^{\lambda+1}} , & \text{if } r_1\geq r_2\\
\end{dcases} }
\]
The continuum hole state wavefunction $\mathrm{\psi_{\infty}}(2)$ corresponding to the ejected electron with orbital angular momentum value $\mathrm{l_d}$ and wave vector $\vec{k}$ is approximated as a plane wave expansion in terms of the spherical Bessel functions and spherical harmonics as:

\begin{equation}
\mathrm{
    \psi_{\infty}(2) = \sum^{\infty}_{l_d = 0}\sum^{l_d}_{m = -l_d}(2l_d+1)i^{l_d}j_{l_d}(kr)Y_{l_d, m}(\theta_2, \phi_2)
}
\end{equation}

The bound state wavefunction were used as defined by Clementi and Roetti \cite{CLEMENTI1974177} using self consistent field (SCF) approximation. Two Slater-type orbitals (STOs) of 1S, 2S and 2P symmetries and three STOs of 3S, and 3P symmetries were used for an accurate estimation of atomic Ar and Ar ion orbital energies \cite{Clementi1963}. The orbital wavefunction $\mathrm{\psi_{i{\lambda_\alpha}} = \sum_pC_{i\lambda_p}\chi_{p\lambda_\alpha}}$, where the basis functions $\mathrm{\chi}$ are Slater-type orbitals with integer quantum numbers \cite{CLEMENTI1974177}. 

The matrix element in equation \ref{Eq4} is further separated using orbital approximation into radial ($\mathrm{D(\lambda,l_d)}$) and angular ($\mathrm{A(m_1,m_1,m_1,\lambda,l_d)}$) overlaps given as follows \cite{gordon1928}:

\begin{widetext}
\begin{equation}
\mathrm{
D(\lambda,l_d) = 4\pi(2l_d+1)(k^{l_d+1/2})^{1/2}\int_0^{\infty}\int_0^{\infty}R_{n_3l_3}^*(r_1)R_{n_2l_2}^*(r_2)\frac{r_{<}^{\lambda}}{r_{>}^{\lambda+1}}R_{n_1l_1}(r_1) j_{l_d}(kr_2)r_1^2r_2^2dr_1dr_2 }
\end{equation}

\begin{equation}
\begin{aligned}
\mathrm{
   A(m_1,m_1,m_1,\lambda,l_d) = \int Y_{1m_2}^*(\theta_1,\phi_1)Y_{1m_3}^*(\theta_2,\phi_2)\sum_{m_\lambda}^{-m_\lambda}
  (-1)^{m_\lambda}Y_{\lambda m_\lambda}(\theta_1,\phi_1)Y_{\lambda -m_\lambda}^*(\theta_2,\phi_2)Y_{1m_1}(\theta_1,\phi_1)}\\
  \times \mathrm{P_{l_{d}}\cos{\theta_2}\sin{\theta_1}\sin{\theta_2}d\theta_1 d\theta_2 d\phi_1 d\phi_2}
  \end{aligned}
\end{equation}
\end{widetext}


Considering the L-MM Auger-Meitner process in Ar, we have calculated the transition probabilities corresponding to three dominant transitions viz $\mathrm{L_{23} - M^2_{23}}$, $\mathrm{L_{23} - M_1M_{23}}$ and $\mathrm{L_{23} - M^2_1}$. The calculated transition rates for atomic Ar and multiply charged $\mathrm{Ar^{q+}}$ ion are given in table \ref{Table3}. The calculated transition probabilities for atomic Ar are in excellent agreement with those reported earlier by Asaad and Melhorn \cite{asaad1968}. 

\begin{table}
\begin{tabular}{|p{0.20\textwidth}|p{0.13\textwidth}|p{0.13\textwidth}|p{0.13\textwidth}|p{0.13\textwidth}|p{0.13\textwidth}| }
\hline
Final hole state & Ar & $\mathrm{Ar^{\cite{asaad1968}}}$ & $\mathrm{Ar^{2+}}$ & $\mathrm{Ar^{3+}}$ & $\mathrm{Ar^{4+}}$ \\ \hline
$\mathrm{3s^{-2}}$ & 0.98 & 1.04 & 2.10 & 2.35 & 4.58 \\ \hline
$\mathrm{3s^{-1}3p^{-1}}$ & 18.5 & 18.0 &34.5 & 47.1 & 61.3 \\ \hline
$\mathrm{3p^{-2}}$ & 46.2 & 53.5 & 83.4 & 99.2 & 117 \\ \hline
\end{tabular}
\caption{Calculated transition probabilities (in a.u.) for atomic Ar and multiply charged $\mathrm{Ar^{q+}}$ ions.}
\label{Table3}
\end{table}

We also note that in estimating $\mathrm{k}$ values, it is important to consider the effect of $\mathrm{2p}$ hole on the binding energies. An initial vacancy in the $\mathrm{2p}$ state results in reduced nuclear shielding. Therefore, the outer shell electrons are more tightly bound in comparison to a filled core shell atom. The effect this reduced shielding is incorporated using the $\mathrm{Z+1}$ or equivalent-core approximation \cite{Sawatzky, Kowalczyk}. In this approximation the final state is treated as if the atom has an effective nuclear charge of $\mathrm{Z+1}$. The electron binding energies are then estimated by using the binding energies of M shells from the next element in the periodic table (potassium in case of argon).

\begin{figure}
    \centering
    \includegraphics[width = 0.9\textwidth]{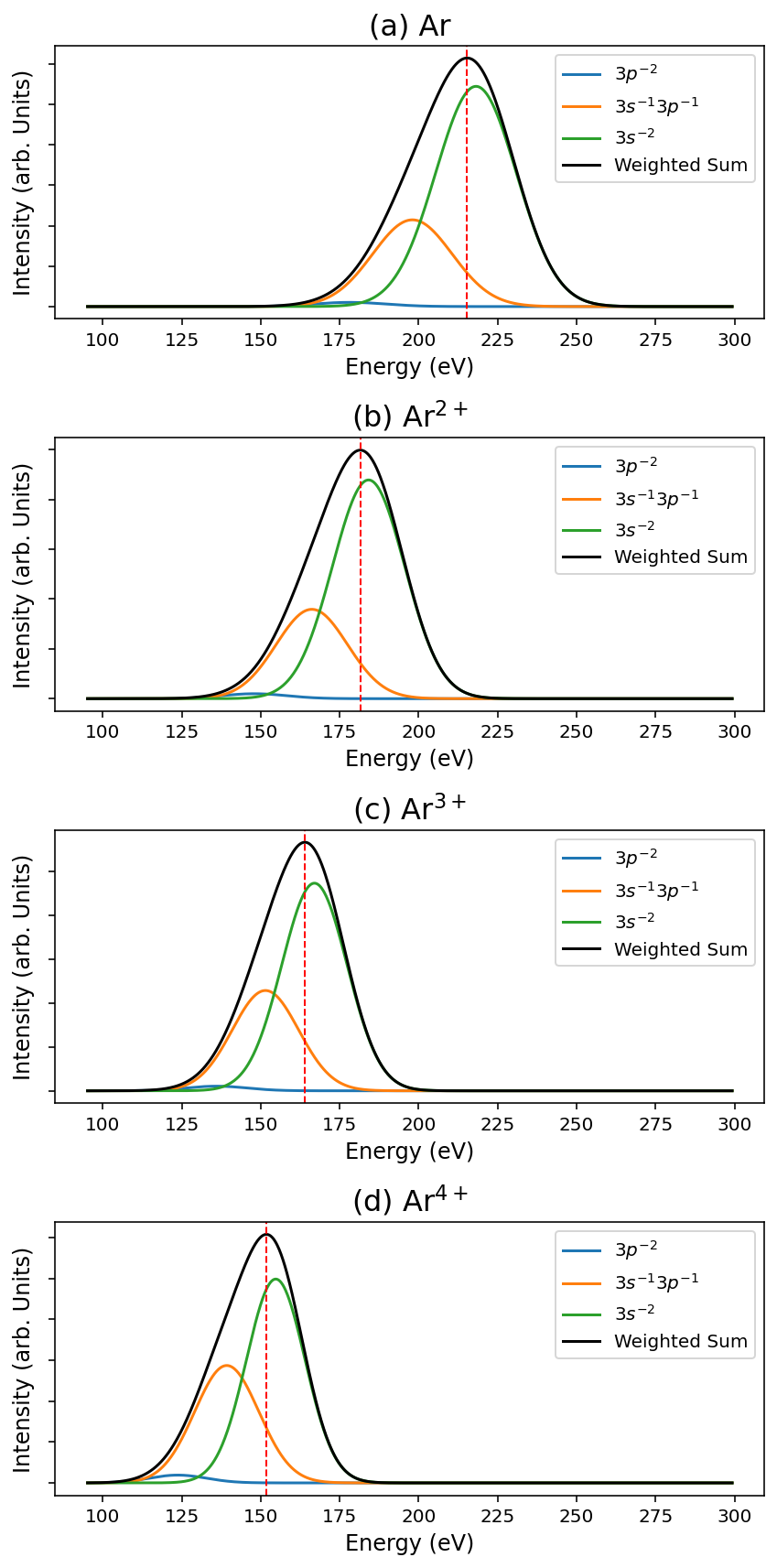}
    \caption{Simulated L-MM Auger-Meitner energy spectra for atomic Ar and multiply charged $\mathrm{Ar^{q+}}$ ions. The red vertical line marks the centroid of the weighted sum spectra.}
    \label{Fig7}
\end{figure}

\subsection{Ar and Ar$^{q+}$ L-MM Auger-Meitner peak energy}

In figure \ref{Fig7} we have plotted the simulated Auger-Meitner spectra for atomic Ar and multiply charged $\mathrm{Ar^{q+}}$ ($\mathrm{q = 2-4}$) ions. The spectra were generated considering the peak energies of the three dominant transitions namely $\mathrm{L_{23} - M^2_{23}}$, $\mathrm{L_{23} - M_1M_{23}}$ and $\mathrm{L_{23} - M^2_1}$. Each transition line was simulated as a Gaussian curve with centroid at the peak energy. The experimental resolution of the spectrometer ($\mathrm{6\%}$) determined the width of the Gaussian. Area under the curves was normalized to the calculated transition probability (see table \ref{Table3}) for respective transitions. The dotted - black curve in figure \ref{Fig7} shows the weighted sum of the three dominant transitions. The peak energy of the cumulative spectra for Atomic Ar is approximately 212 eV and agrees well with the data available in literature. The simulated peak energy is also in good agreement with the experimentally measured value for proton collision with Ar (cf figure \ref{Fig1}). For multiply ionized $\mathrm{Ar^{q+}}$ ions the peak in the simulated spectra shifts towards lower energy as one goes from lower to higher charge state ($\mathrm{Ar^{2+}\rightarrow 182}$ eV; $\mathrm{Ar^{3+}\rightarrow 165}$ eV; $\mathrm{Ar^{4+}\rightarrow 154}$ eV). This is expected due to the increased binding energy of the participating outer shell electrons in ions as compared to neutral atoms. 

We compare the simulated Auger-Meitner spectra in figure \ref{Fig7} with the measured spectra shown in figures \ref{Fig3} and \ref{Fig4} to infer that the measured DDCS peaks for Target and projectile, both, result from the Auger-Meitner decay in multiply charged target ion (instead of single inner shell ionized $\mathrm{Ar^{1+}}$ ion) and projectile ion. The energy peak located at $\sim$ 150 eV refers to Auger-Meitner decay in $\mathrm{Ar^{4+}}$ ion species. Furthermore, equal Auger-Meitner peak energies for the target ion and projectile ion, both, indicate charge state equilibration of the target-projectile ion pair during the initial ionization channel. The equilibrium charge state of the ionized target-projectile ion pair corresponds to $\mathrm{4+}$. We also note that the equilibrium charge state of the target-projectile ion pair is same ($\mathrm{4+}$) for $\mathrm{Ar^{6+} + Ar}$ collision system as well. This signifies the role of multiple ionization and charge exchange processes, possibly leading to quai-molecule formation of the target-projectile ion pair, in low energy collisions.

The low intensity third peak, appearing as a shoulder on the high energy tail in the DDCS spectra (figures \ref{Fig3} and \ref{Fig4}) is attributed to the satellite Auger-Meitner lines \cite{Stolte1974} of multiply charged projectile ions (visible at forward angles) and Auger-Meitner decay from atomic Ar target (cf. figure \ref{Fig1}). At forward angles of measurement the atomic argon Auger-Meitner peak merges with the projectile peak, whereas at backward angles, the projectile satellite peak merges with the prominent target Auger-Meitner peak.

\section{Conclusion}

In summary we have investigated L-MM Auger–Meitner electron emission following collisions of 600 keV $\mathrm{Ar^{3+}}$ and $\mathrm{Ar^{6+}}$ ions with atomic Ar. In this combined experimental and theoretical study, we have measured energy and angle resolved electron emission to reveal distinct contributions from target and projectile Auger–Meitner decay. The measured projectile ion peak energies are in excellent agreement with kinematic calculations considering the Doppler shift in energy in the laboratory frame. We have also measured the The angular SDCS which are well reproduced by an isotropic emission model in the target and projectile rest frame. A theoretical model using the Hartree-Fock SCF wave functions has been developed to  calculate the transition probability for L-shell Auger-Meitner transition in atomic Ar and multiply charged $\mathrm{Ar^{q+}}$ ions. Comparison of the experimentally measured DDCS spectra with the simulated Auger–Meitner spectra for prominent L-MM transitions, reveal that the measured electron emission peaks correspond to multiply ionized $\mathrm{Ar^{4+}}$ ion species rather than singly ionized Ar atoms. In addition, the nearly identical Auger–Meitner peak energies obtained for both $\mathrm{Ar^{3+}}$ and $\mathrm{Ar^{6+}}$ projectile ions signify the important role of multiple ionization, charge exchange, and subsequent charge-state equilibration processes, resulting in an equilibrium charge state of the target–projectile ion pair preceding the Auger-Meitner decay. Our results highlight the significant influence of collision-induced electronic rearrangement on Auger–Meitner decay and provide new insights into the role of multiple vacancy creation, charge equilibration, and quasi-molecular interactions in slow, highly charged ion-atom collisions.

\section{Acknowledgments}
The authors would like to thank Mr. K. V. Thulasiram, Mr. Nilesh Mahtre and Dr. Deepankar Misra for facilitating the operation of the ECR ion accelerator, TIFR, Mumbai.

\section*{Author Contribution}
AHK and RT conceived the study. AHK and LCT performed the experiments. RT and MKH developed the theoretical model. RT analyzed the data. RT and AHK prepared the manuscript.

\section*{Data availability statement}
The data that support the findings of this study are available from the corresponding author upon reasonable request.


\bibliography{apssamp}

\end{document}